# Resonant and non-resonant PL and PLE spectra of CdSe/ZnSe and CdTe/ZnTe self-assembled quantum dots

**T.A. Nguyen, S. Mackowski, L.M. Robinson, H. Rho,
H.E. Jackson and L. M. Smith**
Dept. of Physics, Univ. of Cincinnati, Cincinnati, OH 45221-0011, USA
**M. Dobrowolska, and J.K. Furdyna**
Dept. of Physics, Univ. of Notre Dame, Notre Dame, IN, USA
**G. Karczewski**
Institute of Physics, Polish Academy of Science, Warsaw, Poland

**Abstract.** Using micro- and nano-scale resonantly excited PL and PLE, we study the excitonic structure of CdSe/ZnSe and CdTe/ZnTe self assembled quantum dots (SAQD). Strong resonantly enhanced PL is seen at one to four optic phonon energies below the laser excitation energy. The maximum enhancement is not just one phonon energy above the peak energy distribution of QDs, but rather is 50 meV (for CdSe dots) or 100 meV (for CdTe) above the peak distribution. We interpret this unusual result as from double resonances associated with excited state to ground state energies being commensurate with LO phonons. Such a situation appears to occur only for the high-energy quantum dots.

From extensive work on InGaAs/GaAs self-assembled quantum dots, it is now well known that the electron-phonon interaction is strongly enhanced in quantum confined systems over what is seen in three- and two-dimensions.[1,2] One of the motivations of this work is to explore II-VI self-assembled quantum dots where the electron-phonon interaction in three- and two-dimensions is three orders of magnitude stronger than seen in III-V semiconductors. In this work, we use resonantly excited macro- and nano-PL, along with PLE experiments, to show that the electron-phonon interaction strongly affects the excitation and relaxation processes in CdSe/ZnSe and CdTe/ZnTe self-assembled quantum dots.

The samples were grown by MBE on GaAs substrates. After a thick ZnSe (or ZnTe) buffer layer, several monolayers of CdSe (or CdTe) were put down to form a random distribution of quantum dots. These SAQD layers were then capped by ZnSe (or CdTe). The QDs were excited either non-resonantly by the 514 nm line from an Argon-ion laser, or resonantly by a tunable dye laser (using Rhodamine 590 or Coumarin 540 for CdTe or CdSe dots, respectively). The samples were placed in a helium-vapor cooled cryostat with the temperature ranging from 6 K to 80 K. The emitted PL was dispersed by a DILOR XY triple spectrometer and detected by a LN2-cooled CCD camera.

In Fig. 1 we show non-resonantly excited macro-PL (8 micron laser spot) from CdSe/ZnSe SAQDs along with several resonantly excited spectra for comparison. The PL is strongly enhanced at one to three optic phonon energies below the laser excitation energy. The inset to Fig. 1 shows a detailed view of the enhanced PL which shows several optic phonons associated with ZnSe, CdSe or interface confined LO phonons. If one assumes that the non-resonant PL lineshape reflects the true distribution of QDs, one might expect that the greatest enhancement should occur exactly one LO phonon energy above the maximum intensity of the non-resonantly excited PL. In Fig. 2 we show the variation of the enhanced PL as a function of laser energy, which clearly shows that this simple picture is wrong. The maximum enhancement occurs for laser energies which are 50 meV above the non-resonant PL peak.

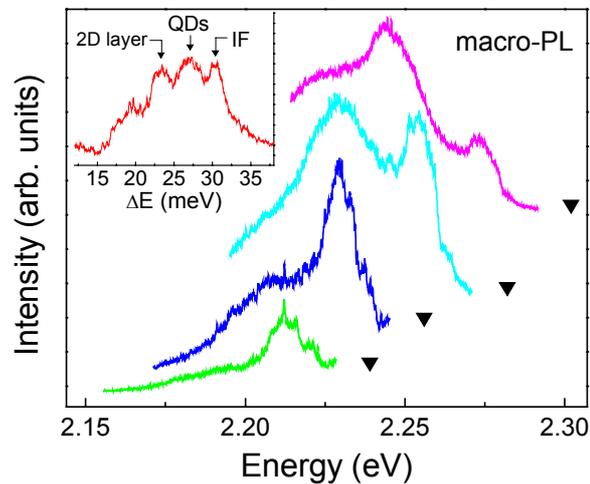

**Figure 1:** Resonant macro-PL of CdSe SAQDs at various excitation energies.

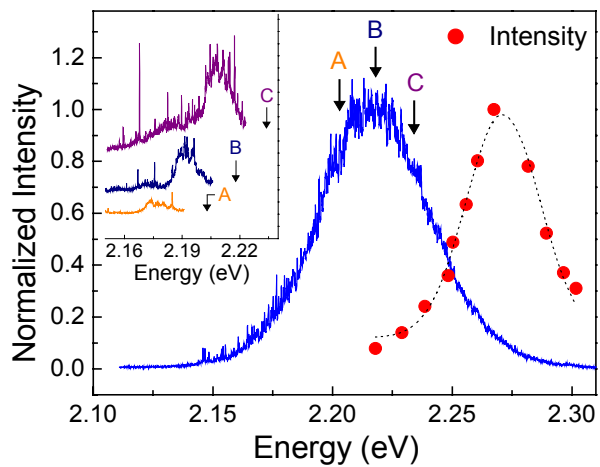

**Figure 2:** Integrated intensity of resonantly excited PL as a function of the excitation energy.

Figure 3 shows similar non-resonant and resonant PL from CdTe/ZnTe SAQDs. In this case, the resonant enhancement clearly continues to increase at energies beyond that seen for the CdSe/ZnSe SAQDs. Figure 4 shows that for the CdTe/ZnTe sample the maximum enhancement does not occur until the laser is ~100 meV above the maximum emission energy. In both the CdSe/ZnSe and CdTe/ZnTe SAQD samples, no evidence for a well-formed wetting layer is seen. A gradually rising continuum of states (presumably associated with barrier states) is seen in detailed PLE measurements. The profile of the enhancement is seen to reflect the distribution of QDs, but shifted by 100 meV.

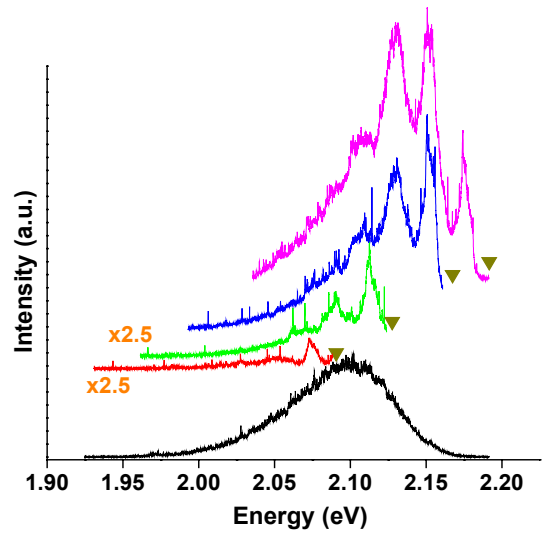

**Figure 3:** Resonant macro-PL of CdTe compared to non-resonant macro-PL (bottom spectrum).

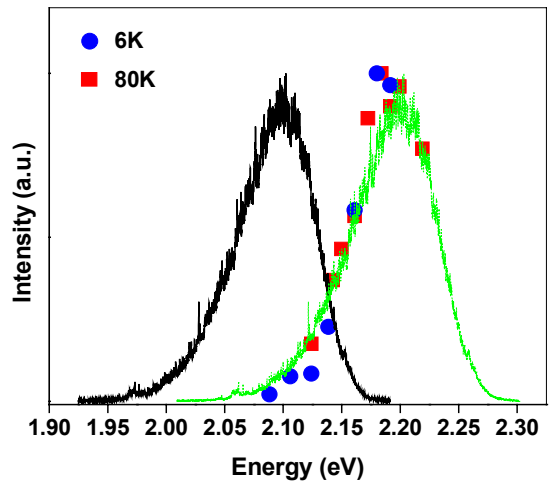

**Figure 4:** Integrated intensity of resonantly excited PL vs. excitation energy at 6K and 80K. The non-resonantly excited PL spectrum shifted by 110 meV is shown underneath the integrated intensity data.

There are several ways to interpret the resonantly enhanced PL. First, the ground states may be directly excited through phonon-assisted absorption. In this case, there is no real confined state at the laser energy. Secondly, the laser may excite directly the ground state of QDs, and the enhanced emission may simply result from phonon-assisted recombination of the ground state. In both these cases, one should expect the processes to be weakly forbidden since they involve (in both cases) a three body processes. The third possibility is that the laser excites directly an excited state of a QD which is efficiently relaxed through LO-phonon emission to the ground state. This would require that the excited state to ground state energy difference be commensurate with an optic phonon. Such a double resonance condition would be rather rare for randomly selected QDs, but when it occurs the enhancement should be very significant. Moreover, this would explain the unexpected 50 and 100 meV shift from the maximum, since perhaps only smaller dots have an excited state to ground state energy difference which is required to support this process.

We have performed some preliminary PLE measurements (not shown here) from nano-scale apertures which seems to show that weak phonon-assisted absorption is present, as well as extremely strong and narrow resonances associated with excited states which are near-resonant with an optic phonon energy. We believe that it is the latter process which explains the rather dramatic phonon-enhanced luminescence.

We gratefully acknowledge the financial support of the National Science Foundation through grants NSF DMR 0071797, 9975655 and 0072897 and the DARPA-SPINS Program.